\documentclass{WileyMSP-template}
\usepackage{color}
\renewcommand{\vec}[1]{\mbox{\boldmath$\mathrm{#1}$}}

\let\sb=_ \catcode`\_=\active \def_#1{\ensuremath \sb{\rm#1}}
\newcommand{\bcorr}{\color{black}}
\newcommand{\bcora}{\color{black}}
\begin{document}

\title{PT-symmetry enabled  spintronic thermal diode and logic gates}

\maketitle


\author{Xi-guang Wang}
\author{Guang-hua Guo}
\author{Jamal Berakdar*}

\begin{affiliations}
Xi-guang Wang, Guang-hua Guo\\
School of Physics and Electronics, Central South University, Changsha 410083, China


Jamal Berakdar\\
Institut f\"ur Physik, Martin-Luther Universit\"at Halle-Wittenberg, 06099 Halle/Saale, Germany\\
Email Address:jamal.berakdar@physik.uni-halle.de

\end{affiliations}


\keywords{Thermal logic gates, unconventional computing, PT-symmetry, spin waves, spin orbit torque, heat flow}

\begin{abstract}

Devices for performing computation and logic operations with low-energy consumption are of key importance for environmentally friendly   data-processing and information technology.  
Here, we present a design for  magnetic elements that use excess heat to perform logic operations. The basic information channel is coupled non-conductive  magnetic stripes with a normal metal spacer. The thermal information signal is embodied in  magnetic excitations and it can be transported, locally enhanced, and controllably steered by virtue of charge current pulses in the spacer. Functionality of essential thermal logic gates is demonstrated by material-specific simulations.  The operation principle takes advantage of the special material architecture with a balanced gain/loss mechanism for magnetic excitation which renders the circuit parity-time symmetric with exceptional points tunable by the  current strength in the spacer. Heat flow at these points can be enhanced, be non-reciprocal, or may  oscillate between the information channels enabling so controlled thermal diode and thermal gate operations. The findings point to a new route for exploiting heat  for useful work on the nanoscale. 

\end{abstract}


\section{Introduction}
The use of excess heat  to perform useful operations is a desirable step in developing environmentally friendly devices. To this end, several concepts have been put forward. For instance, phononic-based  thermal rectifiers 
\cite{PhysRevLett.88.094302},  thermal diodes \cite{PhysRevLett.95.104302}, thermal logic gates \cite{PhysRevLett.99.177208}  were proposed as well as  thermal transistors \cite{doi:10.1063/1.2191730}. Furthermore, nano-thermo-mechanical logic gates, three-terminal magnetic thermal transistors \cite{Castelli2023}, and flexible thermoelectric switches 
\cite{https://doi.org/10.1002/admt.202101043}, and further thermal-circuit elements \cite{doi:10.1126/science.1132898,doi:10.1021/acs.nanolett.2c01100,doi:10.1126/science.abq0883} have been demonstrated experimentally. Considering magnetic materials, the low-energy (thermal) excitations that may serve as information carriers are spin waves (or magnons in reference to their excitation quanta). Generally, the utility of magnons for  information transmission and processing is endorsed by several experiments, e.g.~\cite{Kruglyak2010,Serga2010,Chumak2015,Chumak,Lenk2011}.
Coherent magnon flow and the input/output signals  \cite{Kruglyak2010,Krawczyk2014,Serga2010,Chumak2015,Chumak,Lenk2011,Sadovnikov:2015hm,Wang2018,QWang2020,PhysRevApplied.7.014013,Ren2019,XWang2019} are controllable by several   external parameters such as  magnetic field or spin transfer torques. Material compositions and nano-structuring allow engineering the magnon properties including their amplitude, phase, dispersion and signal attenuation  \cite{Kruglyak2010,Krawczyk2014,Serga2010,Chumak2015,Chumak,Lenk2011,Sadovnikov:2015hm,Wang2018,QWang2020,PhysRevApplied.7.014013,Ren2019,XWang2019,PhysRevB.96.060401,Vogel2018,PhysRevLett.122.197201,PhysRevB.93.020403,doi:10.1063/5.0007095}. \\
Magnons can not only be thermally excited, but they  can also transport heat as a signal to remote heat receptive element \cite{PhysRevB.92.014408,An2013,Wid2016}, which points to the possibility of  magnonic-based thermal signal transfer. Such a possibility is worth investigating  in view of the long diffusion length, negligible Joule  losses (as opposed to charge current pulses), high frequency,  short wavelength (compared to optic-based information processing), and integrability in spintronic devices. Our research presented below indicates that a nonlinear local amplification of  magnonic thermal signal is however necessary for realizing thermal logic gates with sufficient fidelity. 
For this, we exploit a relatively new mechanism for nonlinear signal amplification   based on a magnonic realization \cite{XWang2020,doi:10.1063/5.0029523,XWang2020,doi:10.1063/5.0029523,PhysRevApplied.15.034050}  of PT (parity-time) symmetry \cite{Feng2014,Miri2019,Lin2011,Feng2013,Brandstetter2014,Peng2014natphys,Xu2016,Hodaei2016,Yu2020,Liueaax9144,Galda2019,Zhang2017}.
Typically,     PT-symmetric systems are coupled to their surrounding and exhibit a transition between the  PT-symmetry preserved and broken phases upon changing external parameters. Near the phase-transition (non-Hermitian degeneracy) point, or the exceptional point (EP) the dynamics is highly non-linear with unique properties such as  
excitation  amplification, enhanced sensitivity to external probes, and non-reciprocity.
In our specific case, the magnonic systems are coupled to the surrounding  in such a way that the environment-induced  magnonic  gain and loss are balanced, whilst still realizing a  PT-symmetry setup with EPs. Tuning to  EPs is accomplished by varying the strength of the balanced gain/loss. 
How this situation can be realized experimentally is sketched in Fig.\ref{heatmodel1}:
Two insulating magnonic waveguides (WGs) that may exchange power via RKKY or dipolar interactions serve as
magnonic couplers \cite{Wang2018,QWang2020,PhysRevApplied.7.014013,Ren2019,XWang2019} for steering    signals rendering,  for instance,  directional coupler or magnonic half-adder \cite{Sadovnikov:2015hm,Wang2018,QWang2020,PhysRevApplied.7.014013,Ren2019,XWang2019}. Inserting between the WGs a charge-current carrying metallic layer with strong spin orbit coupling results in PT-symmetric magnonic coupler  with spin-orbit-torque (SOT) magnonic damping of one WG and equally SOT-antidamping in the other WG. The damping/antidamping is caused  by SOT at the interface between the magnetic WG and the normal metal spacer.  This structure possesses EPs tunable by the strength of the magnonic gain/loss effect, which  is linearly dependent on the charge current density in the metal layer. At EPs  the spin dynamics is  highly non-linear \cite{XWang2020}. \\
So far, the heat associated with the spin-wave dynamics  and its possible functionalization in PT-symmetric systems have not been considered. In fact, 
PT symmetry is usually reported for wave systems and it is more delicate  to realize for heat diffusion  \cite{Li170}. {\bcorr We propose and simulate the heat flow embodied in spin waves  that  are triggered and propagate in   magnetic WGs  with sandwiched  non-magnetic metallic spacer with strong spin-orbit coupling. It is shown how an appropriate design   renders the whole structure  PT-symmetric and allows for long-distance  heat diffusion  and control with sufficient fidelity that enables functional thermal logic gates. The existence of EP  is important insofar as  the spin waves amplitude and the induced heat flow are both enhanced there, while the   spin waves can still propagate to a considerable distance and in a  non-reciprocal way,  which is key to realizing   magnonic thermal diodes.}

\section{Theoretical model}
We study two coupled insulating ferromagnetic  waveguides (WG1 and WG2) separated by a charge-carrying spacer with strong spin-orbit coupling, see Fig. \ref{heatmodel1}(a). 
To be specific we use Pt as a spacer layer  in the simulations and assume the WGs to be interlayer-exchange coupled, but the conclusions are also valid  to other spacer materials and WGs coupling schemes.  Yttrium iron garnet (YIG) interfaced with Pt  \cite{https://doi.org/10.1002/admi.202201323,PhysRevB.102.014411,10.1063/1.5025623} is a prototypical example. 
We are interested in magnons that transport heat to a remote distance. The WGs magnetic order as well as the launch and propagation of  magnons   are  governed by the Landau-Lifshitz-Gilbert (LLG) equation \cite{Krivorotov228,Collet2016},
\begin{equation}
\displaystyle \frac{\partial \vec{m}_p}{\partial t} = - \gamma \vec{m}_p \times (\vec{H}_{eff,p} + \vec{H}_{th})+  \alpha \vec{m}_p \times \frac{\partial \vec{m}_p}{\partial t} + \vec{O}_p.
\label{LLG}
\end{equation}
$ p = 1,2 $ enumerates WG1 and WG2. $ M_s $ is the saturation magnetization, and $ \vec{m}_p $ is a unit magnetization  vector field. $ \gamma $ is the gyromagnetic ratio, and $ \alpha $ is the intrinsic (Gilbert) damping constant. The effective field $ \vec{H}_{\mathrm{eff},p} = \frac{2 A_{ex}}{\mu_0 M_s} \nabla^2 \vec{m}_p + \frac{J_{coup}}{\mu_0 M_s t_p} \vec{m}_{p'} + H_B \vec{y} $ includes contributions from the internal exchange field (exchange constant $ A_{ex} $), the interlayer coupling field (coupling constant $ J_{coup} $), and an external magnetic bias field applied along the $ y $ axis, where $ p,p' = 1,2$ and $ p \ne p' $. SOTs   $ \vec{O}_{1(2)} = \gamma c_J \vec{m}_{1(2)} \times(+(-)\vec{y}) \times \vec{m}_{1(2)} $ acting on the two magnetic layers are caused by the charge current in the spacer $\vec{J}_e$, resulting in magnonic loss and gain in WG1 and WG2, respectively.  The effect of SOT is encapsulated in  $ c_J =  \theta_{\mathrm{SH}} \frac{\hbar J_e}{2\mu_0 e t_p M_s} $, which is proportional to  $ J_e $, and the spin Hall angle $ \theta_{SH} $ in the spacer layer.\cite{Collet2016,Krivorotov228} \\
As for thermal effects, we note that it is possible to generate and measure magnons controllably   even at sub-Kelvin temperatures \cite{Demokritov2006,Borisenko2020}.
To capture the influence of  the uncorrelated thermal fluctuations on the spin wave dynamics, one may use thermal random magnetic fields with a correlation function $ \langle H_{th,q}(r,t) H_{th,q'}(r',t') \rangle = \frac{2 \alpha k_B T}{\gamma \mu_0 M_s V} \delta_{qq'} \delta(r - r') \delta(t - t')$ . Here, $ q,q' = x,y,z $, $ k_B $ is the Boltzmann constant, $ T $ is the temperature and $ V $ is the volume of sample. 
The magnetization-dynamics-induced heating is described by the heat diffusion equation,\cite{PhysRevB.92.014408}
\begin{equation}
\displaystyle \frac{dT_p}{dt} = \frac{1}{c_h \rho} \left[k_t \nabla^2T_p + \alpha \mu_0 M_s (\vec{m}_p \times \frac{d\vec{m}_p}{dt}) \cdot \vec{H}_{eff,p} + q_{ext, p}\right],
\label{heat}
\end{equation}
where $ k_t $ is the thermal conductivity, $ c_h $ is the heat capacity, and $ \rho $ is the mass density. The $ \alpha $-dependent  term quantifies   the  heat transfer between spin waves and the lattices in WGs due to  damping   of the collective spin precession in WGs. The Newton term $ q_{ext, p} = c_h \rho (\frac{ T_0 - T_p}{\tau_1} + \frac{T_{Pt} - T_p}{\tau_2}) $ describes the heat exchange between the sample 	at temperature $ T_p $, the surrounding at temperature $ T_0 $, and the spacer material Pt at temperature $ T_{Pt} $, where $ \tau_1 $ and $ \tau_2 $ are the   characteristic thermal relaxation times. The change of temperature $ T_{Pt} $ in the spacer Pt (with charge current density $ J_e $ leading to  heating) is described by,
\begin{equation}
\displaystyle \frac{dT_{Pt}}{dt} = \frac{1}{c_{h,Pt} \rho_{Pt}} \left(k_{t,Pt} \nabla^2T_{Pt} + \frac{J_e^2}{\sigma_{Pt}} + c_{h,Pt} \rho_{Pt}  \frac{ T_1 + T_2 - 2 T_{Pt}}{\tau_2} \right),
\label{heatpt}
\end{equation}
Here, $ \sigma_{Pt} $ is the electrical conductivity of Pt.\\
Solving the three coupled equations (\ref{LLG},\ref{heat},\ref{heatpt}) allows insight into how heat and spin wave excitation  is exchanged between the WGs, spacer, and environment. Generally, experiments with $T_0$ in mK are possible. $T_{Pt}$ depends on $J_e$. As demonstrated in \cite{PhysRevApplied.18.024080}, the required $J_e$ in our setup can be far below the $J_e$ for auto-oscillations.  In the simulations below, we did not optimize the structure as to operate at the lowest $J_e$ that could be achieved, such as by nanostructuring the WGs \cite{PhysRevApplied.18.024080}.\\
For numerical demonstration, we interfaced  Pt  with a Yttrium-Iron-Garnet (YIG), meaning  $ M_s = 1.4 \times 10^5 $ A/m, $ A_{ex} = 3 \times 10^{-12} $ J/m, $ \alpha = 0.005 $, $ J_{coup} = 2.1 \times 10^{-6} $ J/m$ ^2 $. A sufficiently large  magnetic field $ H_B = 1 \times 10^5 $ A/m  drives the system to the remanent state $ \vec{m}_{0,p} = \vec{y} $. The thermal behavior was solved using $ \frac{k_t}{c_h \rho} = 1.5 \times 10^{-6} \rm{m^2\ s^{-1}}, $ $ k_{t} = 8 \rm{W\ m^{-1}\ K^{-1}}, $  $ \tau_1 = 1 \rm{s}, $ and $ \tau_2 = 1 \times 10^{-5} \rm{s} $. For the Pt spacer, parameters $ \frac{k_{t,Pt}}{c_{h,Pt} \rho_{Pt}} = 2.5 \times 10^{-5} \rm{m^2\ s^{-1}}, $ $ k_{t, Pt} = 71.6 \rm{W\ m^{-1}\ K^{-1}}, $ and $ \sigma_{Pt} = 9.43 \times 10^6 \rm{(\Omega m)^{-1}} $. The size of WG1, spacer and WG2 adopted in the simulation is $ 40 \rm{\mu m} \times 4 \rm{nm} \times 5 \rm{\mu m} $ for a $ 4 \rm{n m} \times 4 \rm{nm} \times 5 \rm{\mu m} $ unit simulation cell. {\bcorr Along the spin-wave and heat-diffusion direction (i.e., the $ x $ axis), the cell size is  4 nm,  which is smaller than the exchange length ($ \sqrt{\frac{2A_{ex}}{\mu_0 M_s^2}} =  15.6 $ nm), and hence it is found sufficient   for capturing correctly the spin-wave propagation and related heating effect. Along the $y$-direction (the width), the spin-wave is uniformly excited and its propagation is totally suppressed, and thus a $ 5 {\rm \mu m} $ cell size is adopted. For the numerical implementation we use the Dormand-Prince method (RK45) with a fixed time step of 0.2 ps, for propagating the  equations (\ref{LLG}) and (\ref{heat})   self-consistently.  The temperature distribution enters as  the amplitude of a random field (uncorrelated noise).    We checked, if we  cut the time step for the simulation by half (then 0.1 ps), no visible changes to the results are found, which ensures the convergence of the results.}

\section{Results and Discussion}

\subsection{Magnon heating effect in the PT-symmetric waveguide}
\begin{figure}[htbp]
	\includegraphics[width=0.8\textwidth]{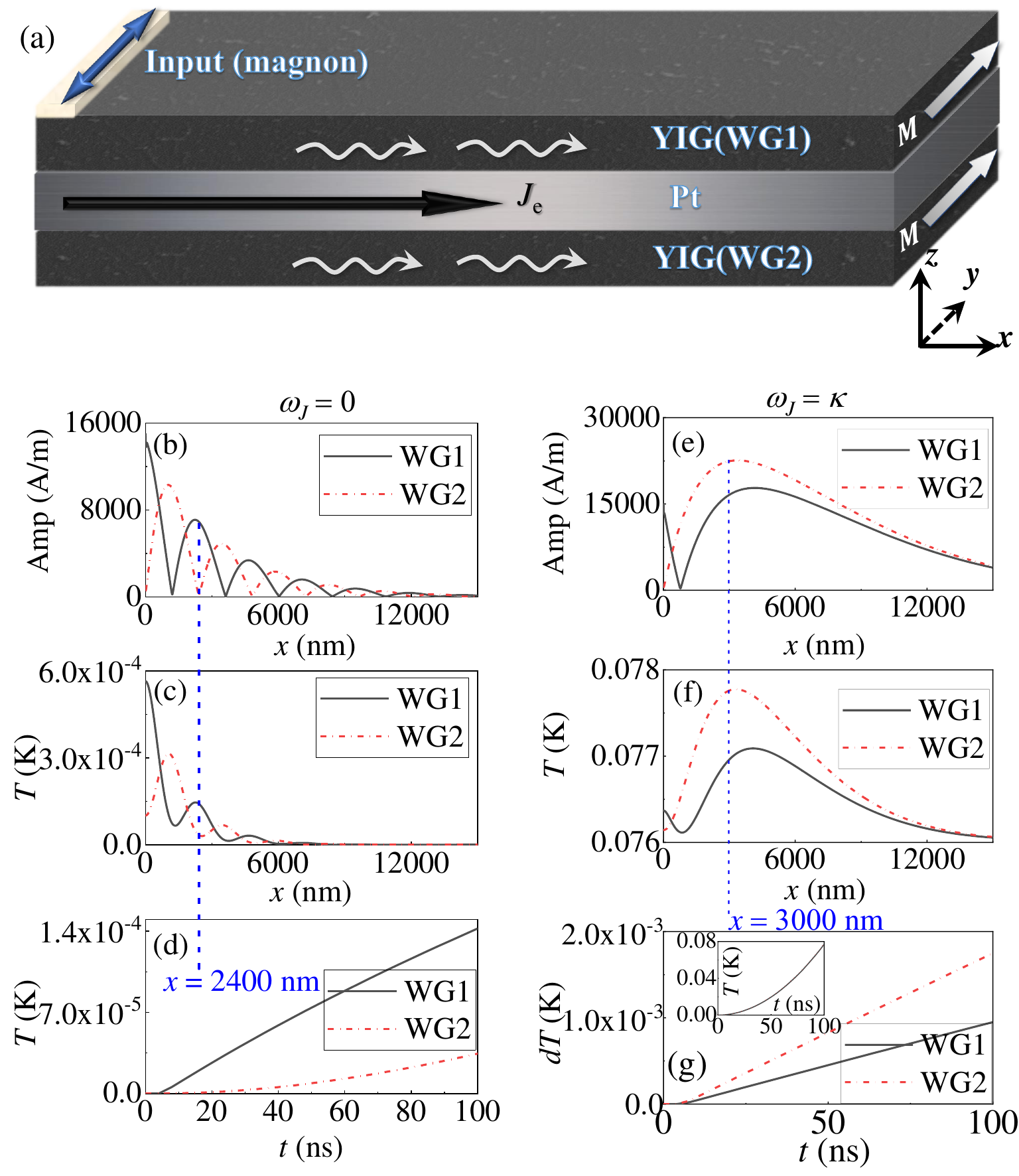}
	\caption{\label{heatmodel1} (a) Magnetic excitation  in two non-conductive magnetic layers serve as planar thermal  waveguides (labeled WG1 and WG2) for transmitting information stored in the temperature of local spin excitations. 
		WG1 and WG2 are magnetically coupled  with a coupling strength $\kappa$. A metallic spacer with a large spin Hall angle is sandwiched between the WGs. Injecting a charge current $ J_e $ in the spacer results in opposite spin-Hall torques on the magnetization in in WG1 and WG2. The torques damp or antidamp the spin excitations  in WG1 and WG2 providing so a balanced gain-loss mechanism (with an effective coupling strength $\omega_J$) and allowing control of the temperature in WG1 and WG2 by varying $ J_e $.  Signals are launched  locally at the left side of WG1 or WG2 and propagate in both layers. (b-d) For $ \omega_J = 0 $.  The signal in WG1 is generated by  a local microwave field $ h_0 sin(\omega_h t) $ ($ h_0 = 75 $ mT and $ \omega_h / (2\pi) = 5 $ GHz) at $ x = 0 $. 
		c) The spin excitation temperature after $100 $ ns. d) Time evolution  of the local spin temperature $ T $ at $ x = 2400 $ nm.
		Switching on  $ J_e $ such that $ \omega_J = \kappa $ changes the (e-g) spatial profiles of the spin excitations  amplitude and the induced temperature. (g) At $ x = 3000 $nm, time evolution of $ T $, and difference $ dT = T - T(h_0 = 0) $ between temperatures with and without the microwave field.	
		In all simulations Joule heating due to $J_e$ is included.}
\end{figure}

We first analyze the eigen-dynamics of magnons, and two eigenfrequencies (optical and acoustic magnon modes) can be obtained as $ \omega_{\pm} = (1-i\alpha)(\omega_0 \pm \sqrt{\kappa^2 - \omega_J^2}) $. Here, we introduce $ \omega_J = \frac{\gamma c_J}{1 + \alpha^2} $, $ \omega_0= \omega_{ex} + \kappa $, $ \omega_{ex} = \frac{\gamma}{1 + \alpha^2}(H_B +\frac{2A_{ex}}{\mu_0 M_s} k_x^2) $ and $ \kappa = \frac{\gamma J_{coup}}{(1+\alpha^2)\mu_0 M_s t_p} $. {\bcorr The  opposing spin-orbit torques acting   on the adjacent magnetic stripes  enhance   the spin-wave damping (meaning more loss in spin-wave amplitude) in one waveguide and decreases the magnons the other waveguide (which means less  loss or  gain in spin-wave amplitude). Such a coupled gain/loss design is found to lead to  antisymmetric imaginary components of the spin-wave potential function, and the whole structure 
	becomes   a PT symmetric system \cite{XWang2020}. Below the gain/loss threshold ($ \omega_J < \kappa $),   the two optic and acoustic modes are separated. At $\omega_J = \kappa$, two modes coalesce (meaning they become  non-Hermitian degenerated),  which is the hallmark of the occurrence of an exceptional point (EP). Above the EP ($\omega_J > \kappa$), the real components of $\omega_{\pm}$ are still the same (with a small $\alpha$), and the two imaginary components are separated.} Without SOT $ \omega_J = 0 $, real parts of $ \omega_{\pm} $ are separated, and two imaginary parts are the same. Exciting two magnon modes with different $ k_x $ and same frequency $ \omega $, because of the interference between the two modes, the spin wave excited at one end of waveguide (say WG1) oscillates between WG1 and WG2, as demonstrated in Fig. \ref{heatmodel1}(b). The induced temperature profile is shown in Fig. \ref{heatmodel1}(c), and a similar oscillating feature is identified. With time, the magnetization dynamics induced temperature is increasing (Fig. \ref{heatmodel1}(d)), and the difference between two guides originates from the different oscillation amplitudes.

\begin{figure}[htbp]
	\includegraphics[width=0.8\textwidth]{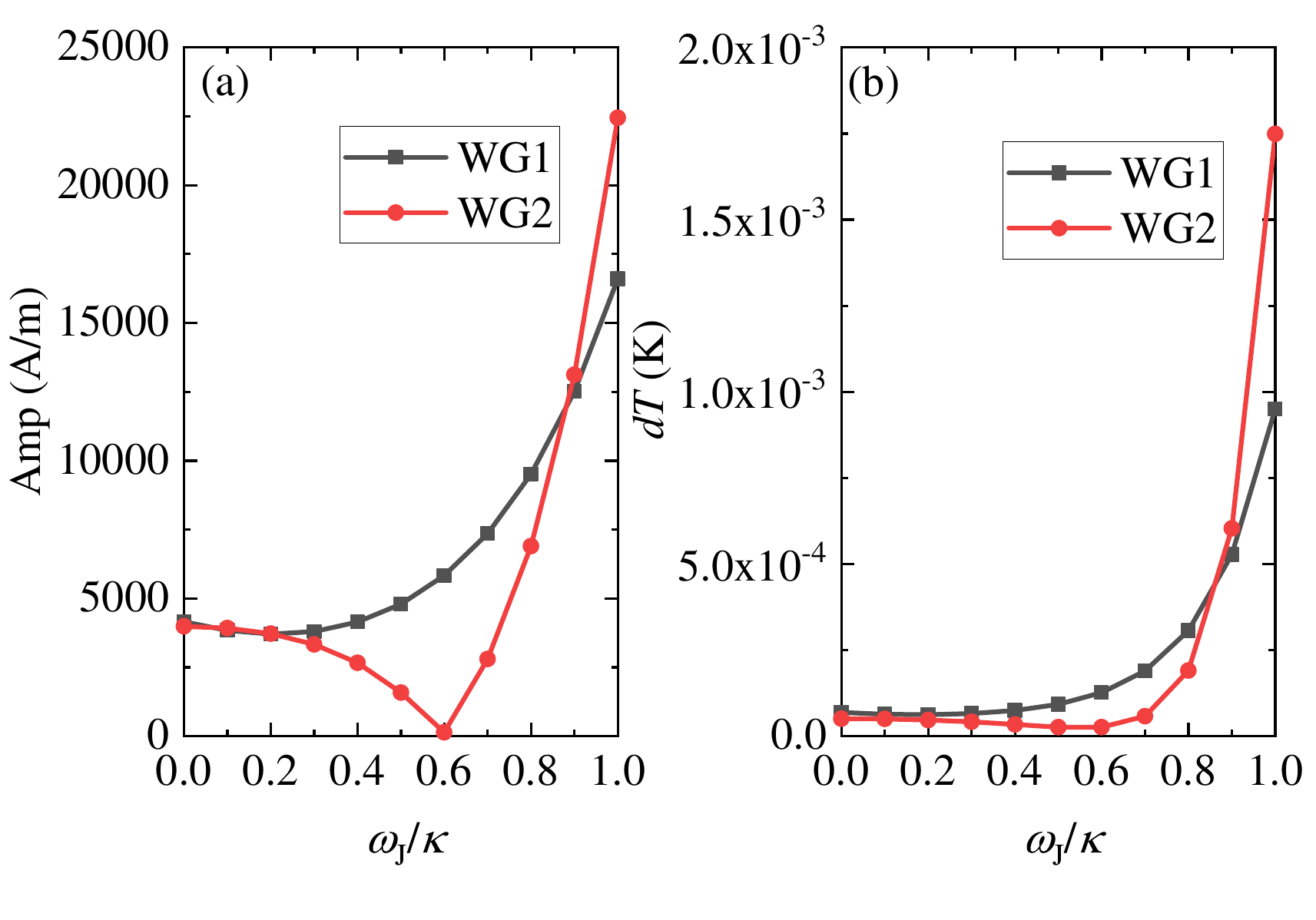}
	\caption{\label{withcj} \bcorr At $ x = 3000 $ nm, the $ \omega_J $ dependent spin-wave amplitude and temperature difference $dT = T - T(h_0 = 0)$ between temperatures with and without the microwave field. The spin-wave is excited by a local microwave field $ h_0 sin(\omega_h t) $ ($ h_0 = 75 $ mT and $ \omega_h / (2\pi) = 5 $ GHz) at $ x = 0 $.}
\end{figure}

A well-defined EP is identified at $ \omega_J = \kappa $ (the corresponding current density $ J_e = 6.4 \times 10^6 {\rm A/cm^2}$), where two eigenfrequencies coalesce at $ \omega_{\pm} = (1 - i \alpha) \omega_0 $ \cite{XWang2020}. The excited spin waves simultaneously propagate in the two WGs, as seen in Fig. \ref{heatmodel1}(e). Compared with the case for $ c_J = 0 $ (Fig. \ref{heatmodel1}(b)), the excited spin wave amplitude   becomes much larger and   propagates through a much longer distance. For the heating effect, the Joule heating in the spacer transfers a uniform temperature to neighboring guides. In  addition to the uniform Joule heating temperature, the propagating spin waves induce an inhomogeneous $ T $ (Fig. \ref{heatmodel1}(f)).  Its value is directly determined by spin wave amplitude. The time-dependent temperatures from Joule heating  and the magnetization dynamics ($ dT = T - T(h_0 = 0) $) are shown in Fig. \ref{heatmodel1}(g). {\bcorr We note that,  due to limited computational resources, our numerical demonstrations   are performed for relatively  small values of the temperature, and hence, the predicted signal is relatively low. In particular, our simulations are  limited to several hundred nanoseconds. Nevertheless, our analysis and results indicate that the spin-wave induced temperature increase is time sustainable. This is important insofar, as in an experimental realization with several milliseconds, a much larger detectable temperature value is to be expected.  On the other side, it should be noted that the operation time of devices based on the scheme presented here will be relatively slow (GHz) but energy-consuming efficient. 

  Another conceptual remark concerns the dynamics  near the EP:  A slight variation in the charge current $\omega_J$  strongly changes the spin-waves and related heating effect, i.e., the existence of EP leads to an  enhanced sensitivity which is crucial for the construction of thermal diodes and gates that operate with sufficient fidelity. An example is seen  in Fig. \ref{withcj}. With the increase in the electric current term $\omega_J$, the nonlinear enhancement in the spin-wave amplitude and $dT$ are evident, especially   around the EP $\omega_J = \kappa$.  At $\omega_J = 0.6\kappa$, the spin-wave amplitude in WG2 is suppressed to zero. This is also related to the $\omega_J$ induced change in the superposition of two magnon modes, and the spin-wave amplitude minimum point is moved to $x = 3000$ nm in WG2 at $\omega_J = 0.6\kappa$.}

\begin{figure}[htbp]
	\includegraphics[width=0.8\textwidth]{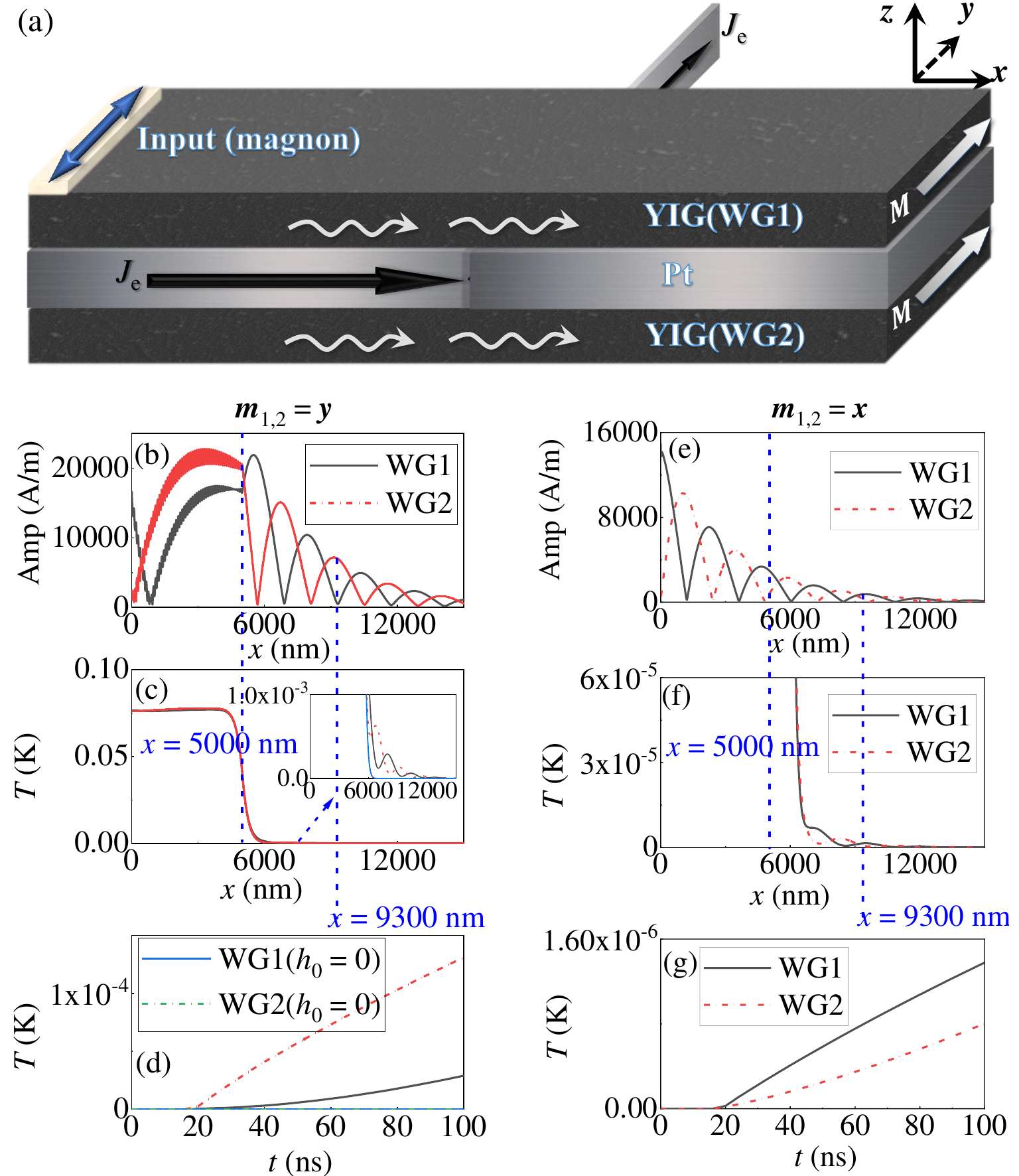}
	\caption{\label{part} (a) The spacer with charge current $ J_e $ is only partly attached to the two magnetic layers ($ x < 5000 $ nm) and the current is diverted out the circuit beyond this region.    For $ \omega_J = \kappa $, (b-c) spatial profiles of magnon amplitudes and induced temperature.  Magnons are launched by the local microwave field $ h_0 sin(\omega_h t) $ ($ h_0 = 75 $ mT and $ \omega_h / (2\pi) = 5 $ GHz) at $ x = 0 $ in WG1. (d) At $ x = 9300 $ nm, time evolution of temperature excited by $ h_0 = 75 $ mT and $ h_0 = 0 $. (e-g) Same curves when the external magnetic field is applied along $ x $. }
\end{figure}

Although  heating  due to magnetization dynamics is much weaker compared to Joule heating, it can transport heat over a very long distance due to the long attenuation length of the spin wave, while Joule heating is localized. To demonstrate this feature, we consider a structure with the Pt spacer partially attached ($ 0  < x < 5000 \rm{nm} $)  to two WGs (Fig. \ref{part}(a)). As shown in Fig. \ref{part}(b), in the range with Pt and SOTs, the excited spin waves still propagate simultaneously in the two guides. Outside this range, the oscillations of spin wave amplitudes are restored. The temperature induced by Joule heating is strongly localized near the Pt spacer, see Fig. \ref{part}(c), while the spin waves can carry   heat over a larger distance. Furthermore, we compare the time-dependent temperatures at $ x = 9300 $ nm for the cases with and without propagating spin waves (Fig. \ref{part}(d)), thus proving this point. We note that the damping constant $ \alpha = 0.005 $ is used in the current simulation. In a realistic case, its value can be much smaller (two orders of magnitude smaller) and the spin wave can propagate over a much larger distance (several micrometers). {\bcorr We note that the influence of the damping constant $ \alpha $ is multifold. On the one hand, the energy transfer between the spin-wave and heat is directly related to the damping term (Eq. (\ref{heat})), and a smaller $\alpha$ decreases the induced temperature amplitude. On the other hand,   the spin-wave amplitude can reach a larger distance, enhancing the remote heating effect.} Furthermore, similar effects can be achieved for significantly lower Joule heating by designing the WGs such that EPs is reached at lower $J_e$ \cite{PhysRevApplied.18.024080}.\\
Changing the direction of the local magnetic field and thus the equilibrium magnetization to $ x $, SOT cannot damp or anti-damp the magnons in this case, and the PT symmetry related phenomena are shut off. This provides a way to manipulate the magnon induced heat flow using the external magnetic field, as demonstrated by Figs. \ref{part}(e-g).  Here, the excited magnon amplitude becomes smaller at $ \omega_J = \kappa $, and the related heating effect becomes obvious weaker.

\subsection{Heat diode design}

\begin{figure}[htbp]
	\includegraphics[width=0.7\textwidth]{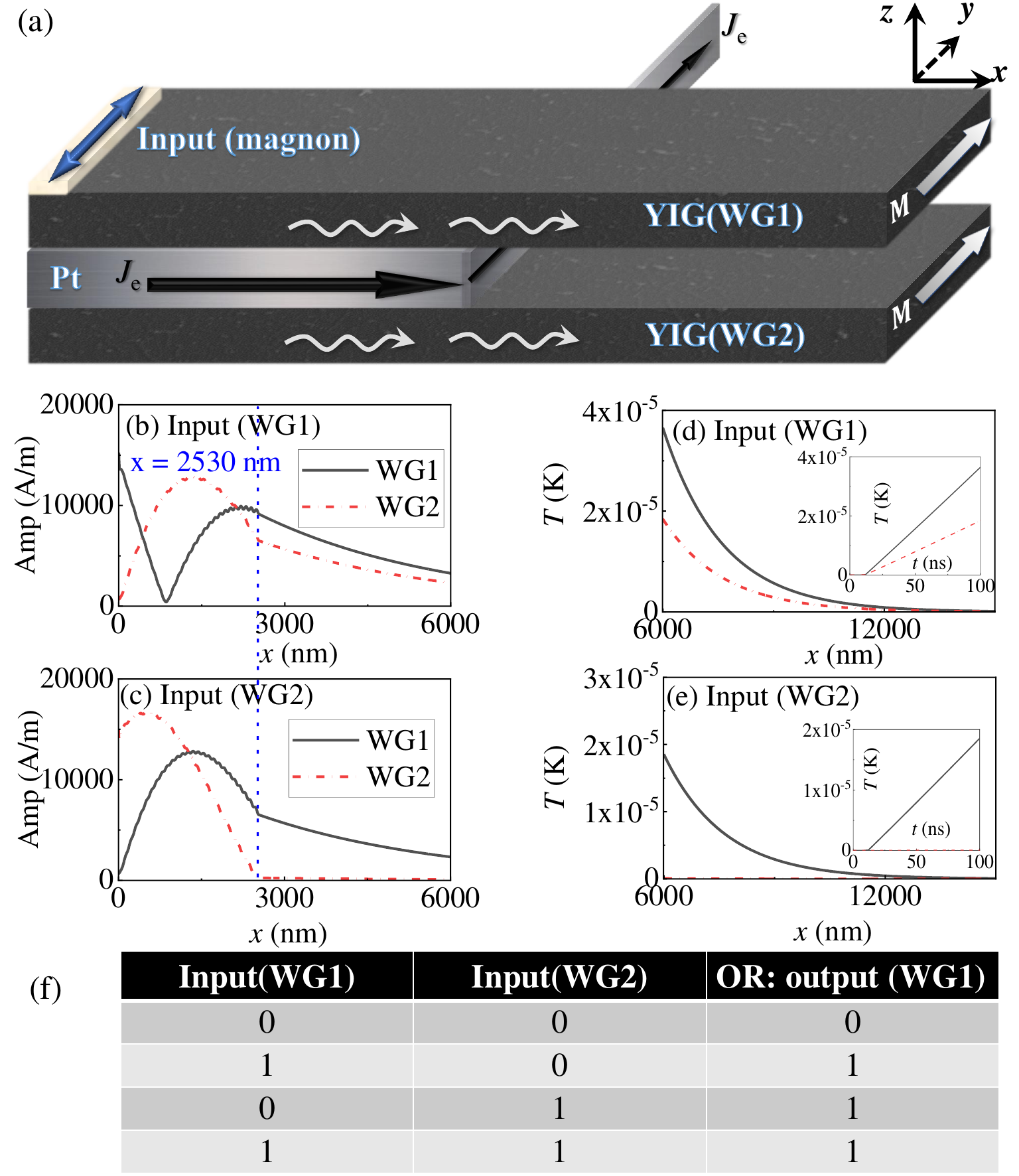}
	\caption{\label{or} (a) The spacer with charge current $ J_e $ only acts on the WGs partially ($ x < 2530 $ nm).   (b-c) For $ \omega_J = 0.7\kappa $, spatial profiles of the magnon amplitudes when the input lies in WG1 or WG2. (d-e) Spatial profiles and time evolution (at $ x = 6000 $ nm) of temperatures heated by magnons in (b) and (c) . (f) Logic "OR" operation table.}
\end{figure}

 Our design for a thermal diode is based on   non-reciprocal spin wave transfer near the EP. The adopted structure is shown in Fig. \ref{or}(a). In the PT symmetric coupled guides, the non-reciprocal spin wave transfer is led by the superposition of two magnon modes, when the charge current $ c_J < \kappa $ is smaller than EP. As demonstrated by Figs. \ref{or}(b-c), the magnon amplitudes emitting from the coupling range with charge current ($ x < 2530 $ nm) become completely different by switching the input (microwave field) from WG1 to WG2. Especially, for input in WG2, the emitting magnons are mainly located in WG1, while the amplitude in WG2 approaches 0 which leads to  a non-reciprocal heat flow, i.e.,~a realization of a thermal diode, as demonstrated by Figs. \ref{or}(d-e). The time-dependent temperature $ T $ in shown in the insets of Figs. \ref{or}(d-e),   exhibiting  the characteristic features of the thermal diode.

Based on the thermal diode above, we design several possible logic operations.  The logic "OR" operation can be realized in the heat diode structure, as demonstrated by Fig. \ref{or}(f). Here, the input (magnon excited by microwave field) in WG1 or WG2 is treated as logic "1", and the nonzero output magnon amplitude in WG1 represents logic "1". Only in the case without any input in WG1 and WG2, the logic output in WG1 is "0". Otherwise, the output is always logic $ 1 $, meaning a logic "OR" operation between the two inputs (WG1 and WG2).

\begin{figure}[htbp]
	\includegraphics[width=0.8\textwidth]{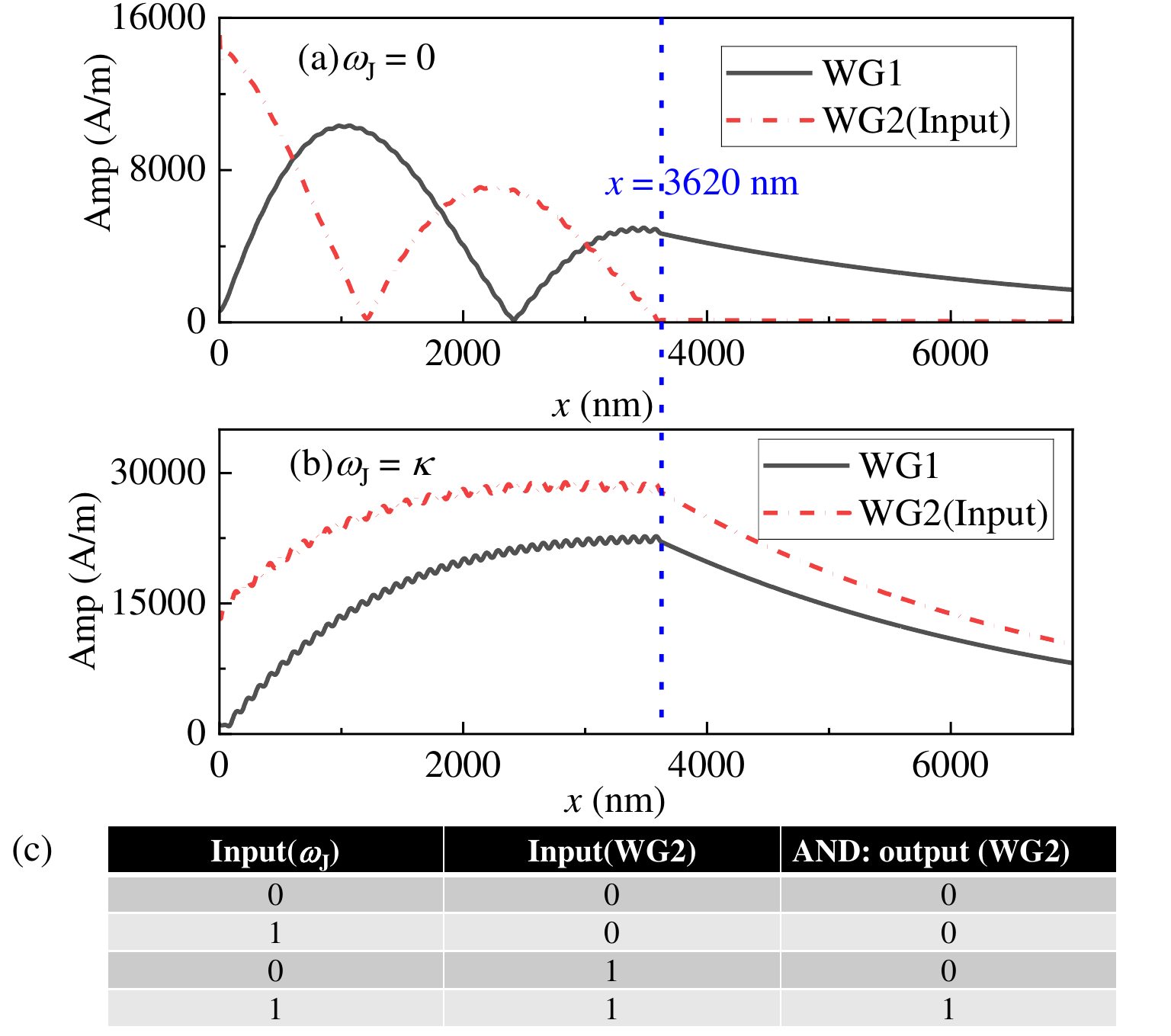}
	\caption{\label{and} Applying the local microwave field $ h_0 sin(\omega_h t) $ ($ h_0 = 75 $ mT and $ \omega_h / (2\pi) = 5 $ GHz) at $ x = 0 $ in WG2, spatial profiles of magnons for (a) $ \omega_J = 0 $ and (b) $ \omega_J = \kappa $. (c)Logic "AND" operation table based on the results in (a-b).  Here, the range with SOTs and interlayer coupling is $ x < 3620 $ nm. }
\end{figure}

\begin{figure}[htbp]
	\includegraphics[width=0.8\textwidth]{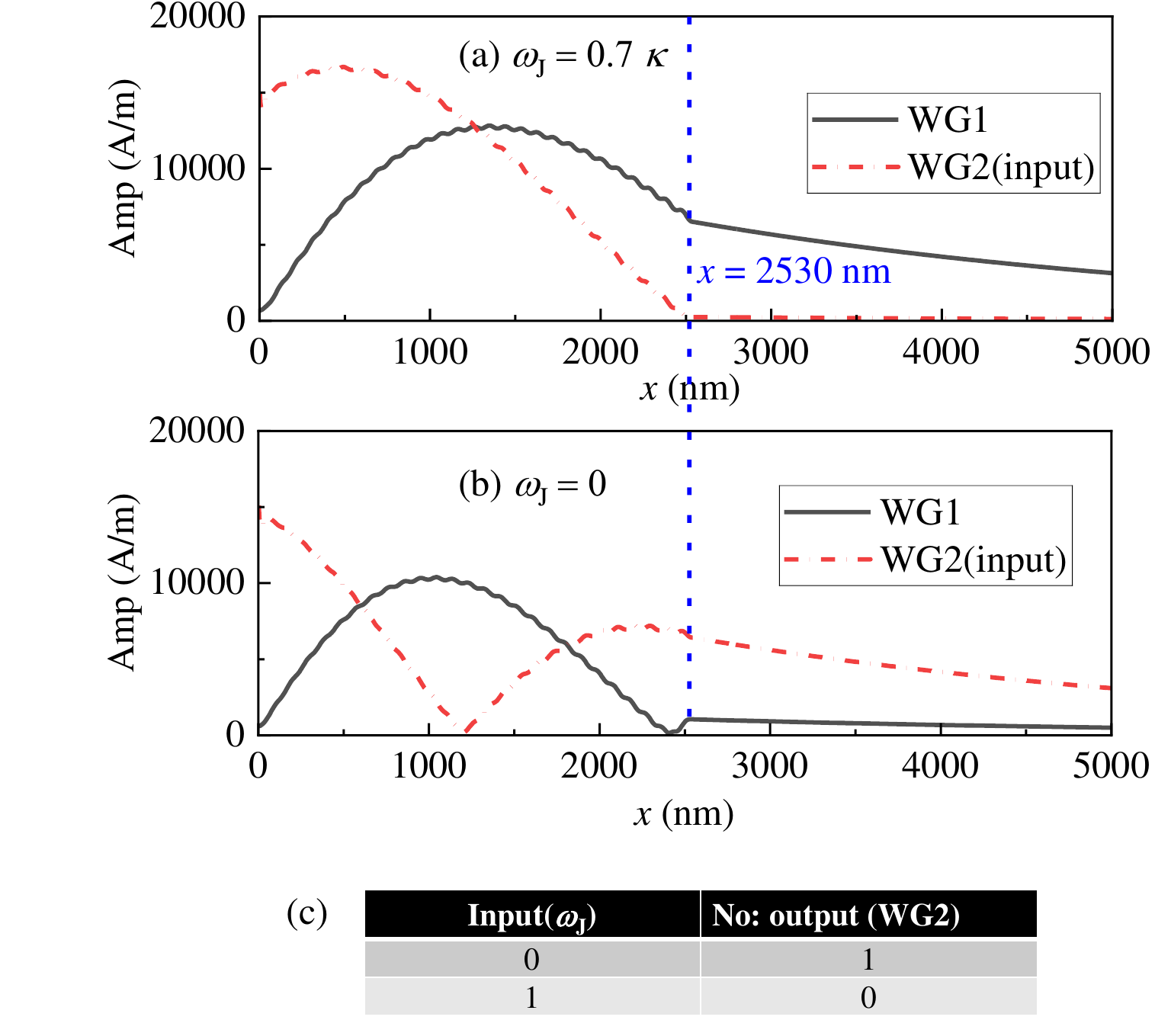}
	\caption{\label{no} Spatial profiles of magnons for (a) $ \omega_J = 0.7 \kappa $ and (b) $ \omega_J = 0 $. The range with SOTs and interlayer coupling is $ x < 2530 $ nm. (c) Logic "NO" operation table. }
\end{figure}

The logic "AND" operation   can be realized in a similar structure, as demonstrated in Fig. \ref{and}(a-c). Here, $ \omega_J = \kappa $ and input in WG2 are treated as logic "1", and the nonzero output magnon amplitude in WG2 is logic "1".  Only for $ \omega_J = \kappa $ with input in WG2, the logic output in WG2 is "1", meaning a logic "AND" operation is realized. The logic "NO" operation is demonstrated in Fig. \ref{no}.  Electric current input $ \omega_J = 0.7 \kappa $ is logic "1". Keeping the magnon input in WG2, and the output in WG2 is "0" when  $ \omega_J = 0.7 \kappa $. Switching off the electric current, the output becomes "1". Noteworthy, for the above operations based on PT-symmetric structure, by varying the amplitude of the charge current, we can further dynamically switch between these logic functions.

\begin{figure}[htbp]
	\includegraphics[width=0.8\textwidth]{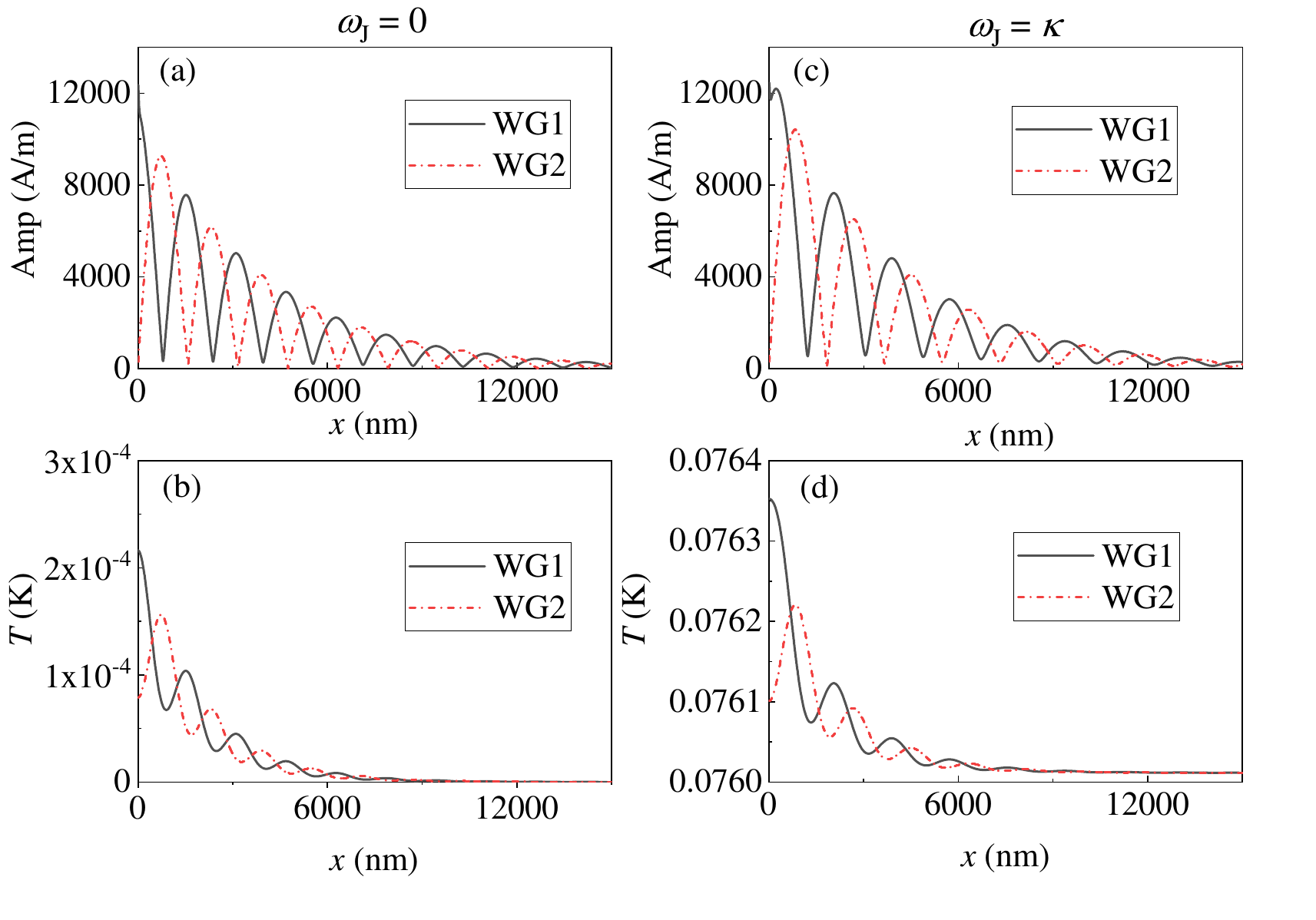}
	\caption{\label{dipolarwhole} \bcora Simulation for the model Fig. \ref{heatmodel1}(a) with dipolar interaction. For $ \omega_J = 0 $ (a-b) and $ \omega_J = \kappa $ (c-d), the spatial profiles of spin-wave amplitude and induced temperature.  The signal in WG1 is generated by  a local microwave field $ h_0 sin(\omega_h t) $ ($ h_0 = 75 $ mT and $ \omega_h / (2\pi) = 5 $ GHz) at $ x = 0 $. The spacer with charge current $ J_e $ acts on the whole WGs.}
\end{figure}

\begin{figure}[htbp]
	\includegraphics[width=0.8\textwidth]{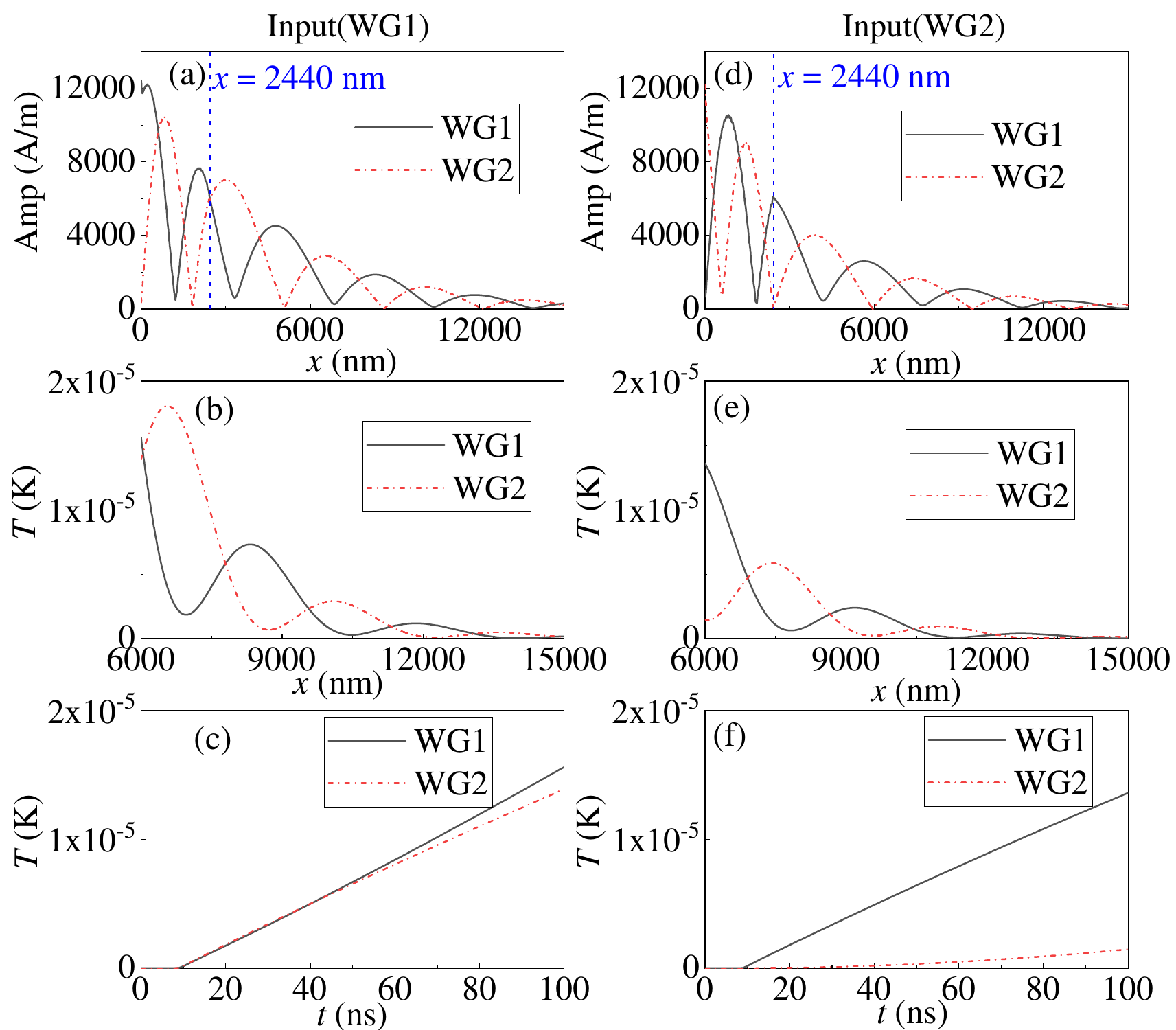}
	\caption{\label{dipolarpart} \bcora Simulation for the model Fig. \ref{or}(a) with dipolar interaction. When the input lines in WG1 (a-c) or WG2 (d-f),  the magnon amplitudes profile(a,d), temperature profile (b, e), and time-dependent temperature at $x = 6000$ nm (c, f). The spacer with charge current $ J_e $ ($ \omega_J = \kappa$) acts on the WGs partially ($ x < 2440 $ nm).}
\end{figure}

\bcora Besides, to be more close to experimental realization, we consider the influence of dipolar interaction. We perform the simulations for the coupled nanostripes (the size is still $40 {\rm \mu m} \times 4 {\rm nm} \times 5 {\rm \mu m} $) and include the dipolar interaction field $\vec{H}_{\mathrm{demag}}$,
	\begin{equation}
	\displaystyle \vec{H}_{\mathrm{demag}}(\vec{r}) = -\frac{M_s}{4 \pi} \int_{V} \nabla \nabla'\frac{1}{|\vec{r} - \vec{r'}|}\vec{m}(\vec{r'})d\vec{r}' .
	\label{demag}
	\end{equation} 
Still, we observe the similar oscillating behavior for the case without electric current $\omega_J = 0$ in Fig. \ref{dipolarwhole}(a-b). Due to that the dipolar interaction increases the coupling between WG1 and WG2, at $\omega_J = \kappa$ (the exceptional point (EP) for the case without dipolar interaction), there are still SW amplitude oscillation as the EP ($ \omega_J > \kappa $) is increased by the dipolar interaction (Fig. \ref{dipolarwhole}(c-d)). Here the spin-wave transfer is non-reciprocal, and one can realize the similar design of the diode and logic gates. As an example, in Fig. \ref{dipolarpart}, when the charge current $\omega_J = \kappa$ is locally applied in the range of $x < 2440 $ nm (same to the model of Fig. 4(a)), we obtain different spin-wave amplitudes outside this range. We note due to the dipolar interaction, the SW is still periodically transferred between two WGs in the range without interlayer coupling ($J_{coup}(x > 2440 {\rm nm}) = 0$). Switching the input from WG1 to WG2, the temperature $T$ near the position $x = 6 {\rm \mu m}$ (i.e. output) becomes completely different(Fig. \ref{dipolarpart}(c, f)), enabling the possibilities of thermal diode and logic "OR" operation similar to the findings in Fig. \ref{or}.

\color{black}	

\subsection{Electric current pulse and {\bcorr local} temperature gradient}

\begin{figure}[htbp]
	\includegraphics[width=0.8\textwidth]{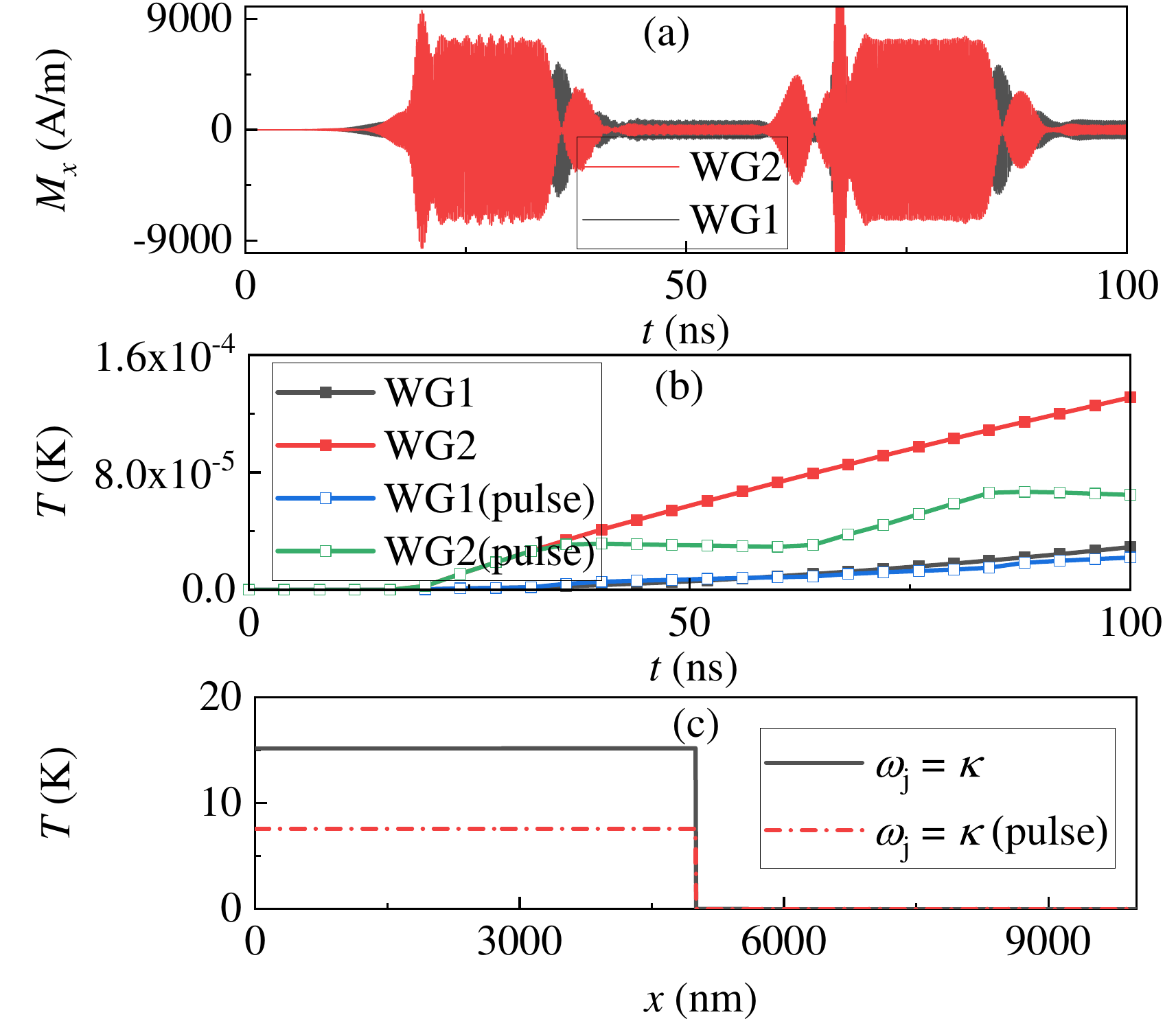}
	\caption{\label{pulset} Driven by continuous electric current and pulsed electric current with $ \omega_J = \kappa $,  (a) time dependent $ M_x $ detected at $ x = 9300 $ nm in WG1 {\bcorr and WG2}. (b)At $ x = 9300 $ nm, time dependent temperature $ T $.  (c) Temperature profile in the spacer between WG1 and WG2 at time $ t = 100 $ ns.}
\end{figure}

To decrease the overall energy consumption of the logic operations,  we apply a charge current pulse $J_e(t)$. For a sequence of charge currents with a period of 50 ns  (Fig. \ref{pulset}), the Joule heating induced temperature is decreased by factor 2, compared to DC $J_e$.  When the charge current is switched on, magnons at EP are enhanced (Fig. \ref{pulset}(a)), and the enhancement disappears soon when the current is switched off. It also affects the magnon induced heating in two guides, Under the pulsed current, the heating effect becomes obvious weaker compared to that for continuous current (Fig. \ref{pulset}(b)). The enhanced  nonlinearity of the magnetization  dynamics at EPs is favorable for speeding up the logic gate operations. The degree of nonlinearity can be increased by having higher-order EPs which can be achievable through WGs design, as demonstrated in Ref.\cite{PhysRevApplied.15.034050}.

\begin{figure}[htbp]
	\includegraphics[width=0.5\textwidth]{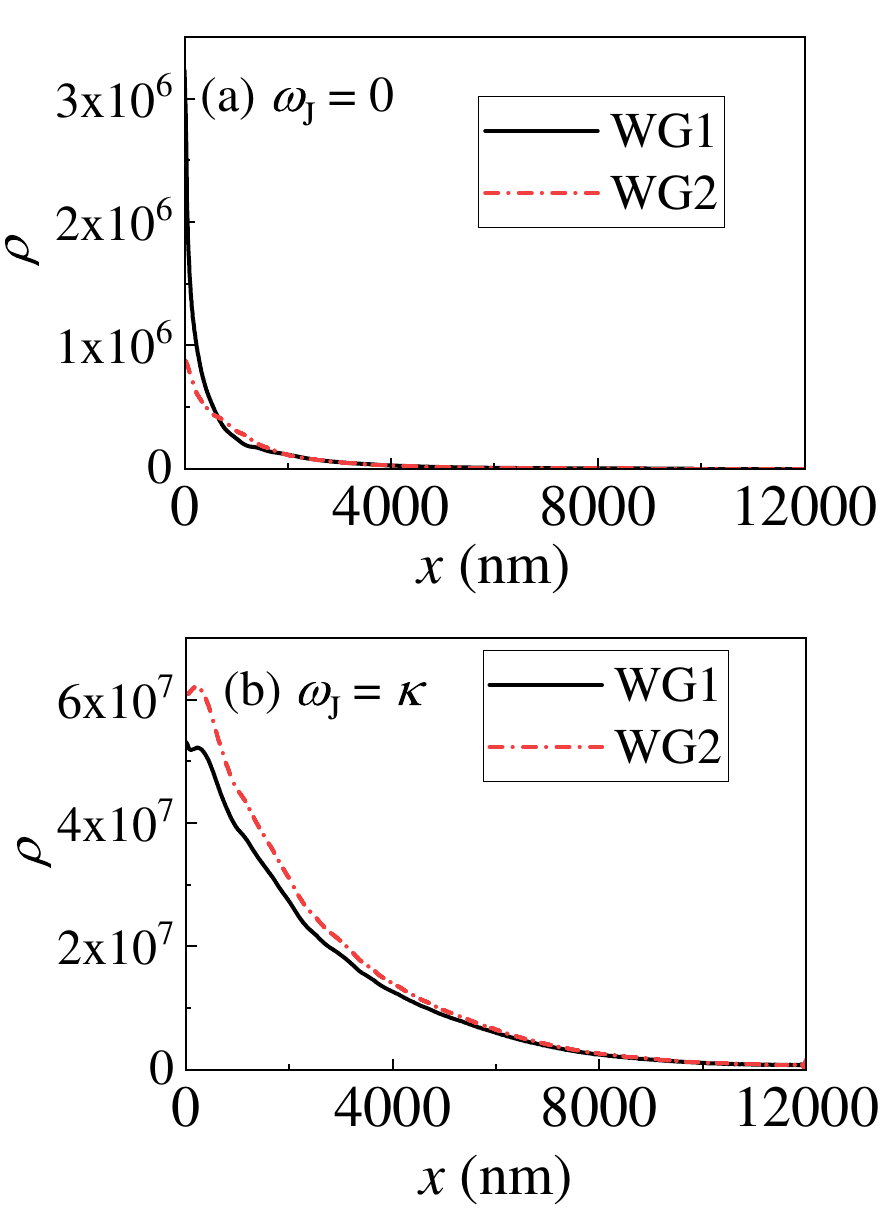}
	\caption{\label{xrhof} With (a) $ \omega_J = 0 $ and (b) $ \omega_J = \kappa $, spatial profile of the magnon density $ \rho = M_x^2 + M_z^2 $. Here, a local temperature $ 100 $ K is set at the left end in WG1 ($ x < 40 $ nm). }
\end{figure}

The propagating magnons  can be excited via the temperature gradient. Here, we set a local temperature $ T = 100 $ K at the left end, and thermally excited magnons propagate towards the rest part of the sample. As demonstrated in Fig. \ref{xrhof}(a), thermal magnons can transmit into neighboring WGs via coupling . As thermal magnon frequencies lie in a wide range and the magnon beating length changes with frequency, we do not observe a clear periodic energy transfer between the two guides. Applying $ \omega_J = \kappa $, magnons with all frequencies reach EP simultaneously, and there is still an obvious magnon enhancement in the two guides, see Fig. \ref{xrhof}(b).

\subsection{Relaying magnon propagation}

\begin{figure}[htbp]
	\includegraphics[width=0.92\textwidth]{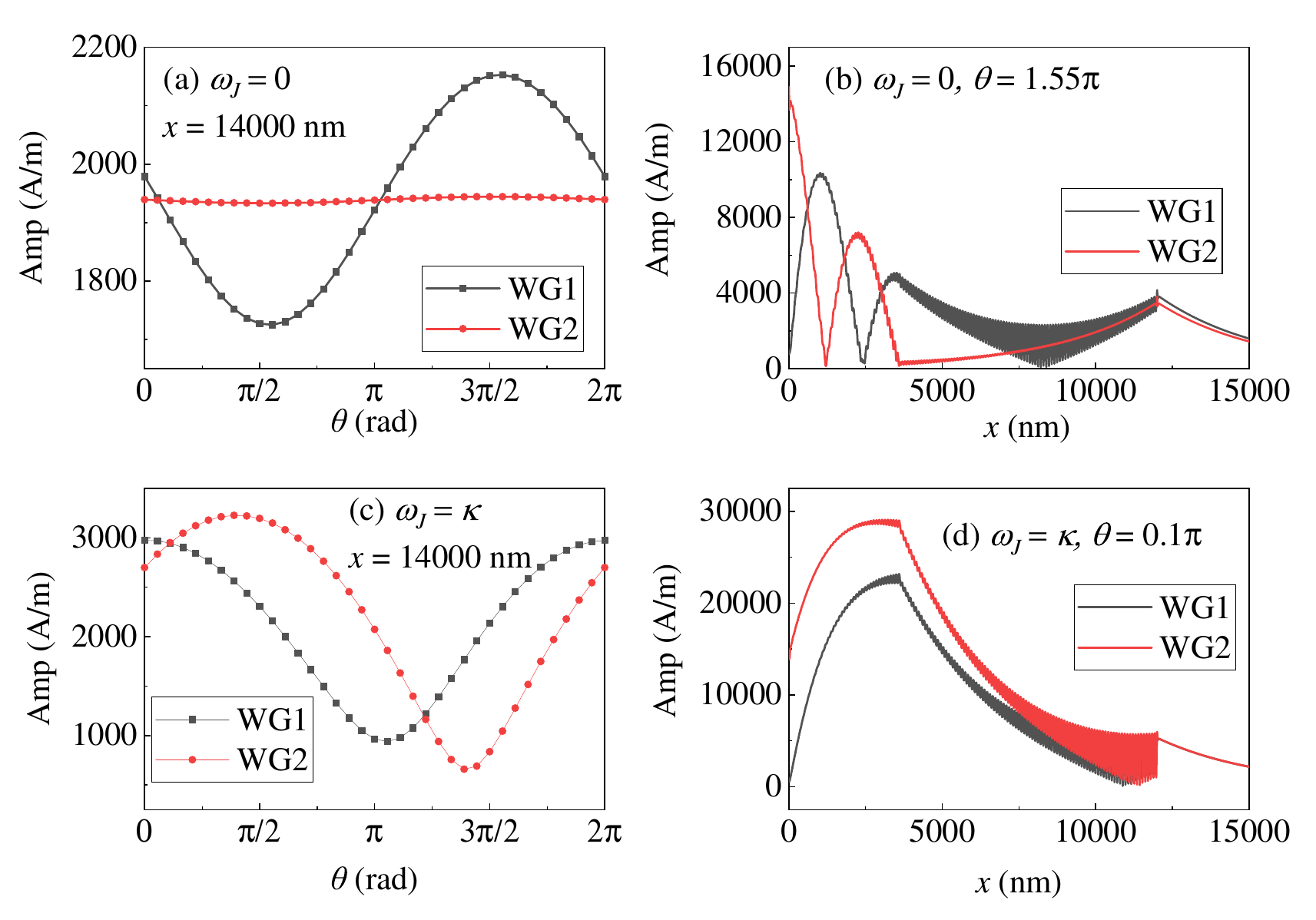}
	\caption{\label{relay} Magnons are injected via applying the local microwave field $ h_0 sin(\omega_h t) $ ($ h_0 = 75 $ mT and $ \omega_h / (2\pi) = 5 $ GHz) at $ x = 0 $ in WG2. At $ x = 12000 $ nm, a second microwave field $ h_1 sin(\omega_h t + \theta) $ ($ h_1 = 37.5 $ mT) is applied to both guides to relay the injected magnons. (a,c) The phase $ \theta $ dependence of magnon amplitude at $ x = 14000 $ nm when $ \omega_J = 0 $ or $ \kappa $. (b,d) The spatial profiles of magnon amplitudes. Here, the range with SOTs and interlayer coupling is $ x < 3620 $ nm.}
\end{figure}

To relay the propagating magnons in the WGs, we apply a second microwave field to both WGs. As shown in Fig. \ref{relay}, when $ \omega_J = 0 $, in WG1 the magnon amplitude changes with the phase angle $ \theta $ of the second microwave field, while in WG2 it remains unaffected. Here, the range with interlayer coupling is $ x < 3620 $ nm, and without the second microwave field, the emitting magnons mainly distribute in WG1 (Fig. \ref{and}(a)). At $ x = 14000 $ nm, the amplitude in WG1 is $ 200 $ A/m. With the magnons excited by the second microwave field, the interference between two magnon currents induces the $ \theta $ dependence amplitude. For $ \theta = 1.55 \pi $, the larger amplitude in WG1 is kept, and the absolute values of magnon amplitudes in both WG1 and WG2 are increased. Furthermore, when $ \omega_J = \kappa $, the emitted magnons distribute over the two WGs, and the second microwave field can equally enhance the magnon amplitude in WG1 and WG2 for $ \theta = 0.1 \pi $. Here,   interference can affect the magnon amplitudes of the two WGs.

\section{Conclusion}

The medium for the thermal information signals is  magnons  in coupled PT-symmetric non-conductive magnetic stripes. PT-symmetry  is essential to have an EP (or non-Hermitian degeneracy point) at which the information signals can be amplified or steered controllably.  The PT-symmetry is brought out by  a balanced gain and loss of magnonic excitations due to SOTs at the interface between the spacer and magnetic stripes. A charge current pulse sets the value of the SOT. The current-induced Joule heating is strongly localized.  The heat flow associated with magnons can propagate through  a long distance (compared to phononic thermal flow), allowing for remote thermal information exchange.   The charge current  strength tunes the device to the  non-reciprocal heat flow regime, which is necessary for thermal diodes enabling several types of reconfigurable logic operations. The results demonstrate an example for PT symmetry in   heat diffusion and  point to new designs of heat devices.

\medskip
\textbf{Acknowledgments} \par 
This work was supported by the DFG through SFB TRR227, and Project Nr. 465098690, the National Natural Science
Foundation of China (Grants No. 12174452, No. 12074437, No. 11704415, and No. 11674400), and the Natural Science Foundation of Hunan Province of China (Grants No. 2022JJ20050, No. 2021JJ30784, and No. 2020JJ4104), and the Central South University Innovation-Driven Research Programme (Grant No. 2023CXQD036).

\medskip

%
\bibliographystyle{MSP}
\bibliography{achemso-demo}


\end{document}